\documentclass[manuscript]{aastex}
\usepackage{lineno}
\usepackage{bm}
\usepackage{color}
\usepackage{esint}
\usepackage{appendix}
\usepackage{multirow}
\usepackage{makecell}
\usepackage{threeparttable}
\usepackage{url}
\usepackage{ragged2e}
\usepackage{longtable}
\usepackage{rotating}
\usepackage{footnote}
\usepackage{diagbox}

\begin{document}
\title{Widespread 1--2 MeV Energetic Particles Associated with Slow and Narrow Coronal Mass Ejections: Parker Solar Probe and STEREO Measurements}
\author{Bin Zhuang\altaffilmark{1}, No\'{e} Lugaz\altaffilmark{1}, and David Lario\altaffilmark{2}}
	
\affil{$^1$Institute for the Study of Earth, Ocean, and Space, University of New Hampshire, Durham, NH, USA; \url{bin.zhuang@unh.edu}, \url{noe.lugaz@unh.edu} \\
	$^2$NASA, Goddard Space Flight Center, Heliophysics Science Division, Greenbelt, MD 20771, USA; \url{david.larioloyo@nasa.gov} \\} 

\begin{abstract}
Suprathermal ions in the corona are thought to serve as seed particles for large gradual solar energetic particle (SEP) events associated with fast and wide coronal mass ejections (CMEs). A better understanding of the role of suprathermal particles as seed populations for SEP events can be made by using observations close to the Sun. We study a series of SEP events observed by the Integrated Science Investigation of the Sun (IS$\odot$IS) suite on board the Parker Solar Probe (PSP) from 2020 May 27 to June 2, during which PSP was at heliocentric distances between $\sim$0.4 and $\sim$0.2 au. These events were also observed by the Ahead Solar TErrestrial RElations Observatory (STEREO-A) near 1 au, but the particle intensity magnitude was much lower than that at PSP. We find that the SEPs should have spread in space as their source regions were distant from the nominal magnetic footpoints of both spacecraft, and the parent CMEs were slow and narrow. We study the decay phase of the SEP events in the $\sim$1--2 MeV proton energy range at PSP and STEREO-A, and discuss their properties in terms of both continuous injections by successive solar eruptions and the distances where the measurements were made. This study indicates that seed particles can be continuously generated by eruptions associated with slow and narrow CMEs, spread over a large part of the inner heliosphere, and remain there for tens of hours, even if minimal particle intensity enhancements were measured near 1 au.
\end{abstract}

\justifying
\section{Introduction}
The two-class paradigm for solar energetic particle (SEP) events \citep[e.g.,][]{1995RvGeo..33S.585R} suggests that the acceleration of SEPs near the Sun occurs in connection with (1) magnetic reconnection processes in solar flares resulting in impulsive events, and (2) shock waves generated by fast and wide coronal mass ejections (CMEs) resulting in gradual events. Large gradual SEP events (as per NOAA definition those SEP events where $>$10 MeV proton intensities exceed 10 pfu; $1 \ \rm{pfu}=1 \ \rm{proton} \ \rm{cm}^{-2} \ \rm{s}^{-1} \ \rm{sr}^{-1}$) are of particular interest because of their extreme space weather effect \citep[e.g.,][]{2010SpWea...8.0E09C,2020SW002665}. Although diffusive shock acceleration \citep[DSA;][]{1983JGR....88.6109L,2012SSRv..173..247L} is widely accepted as the main mechanism for particle acceleration at shocks in large gradual SEP events, the processes involved in the acceleration of particles are still a major open question in heliospheric physics \citep[see the introduction in][]{2015PhR...557....1V}. In particular, a fundamental problem is determining the seed particle populations on which shocks act \citep{2015PhR...557....1V}.

Several studies have suggested a connection between the suprathermal particles measured in situ near 1 au and the seed particles over which the shocks expanding out from the Sun act \citep{2005AIPC..781..219M,2006ApJ...649..470D,2012AIPC.1436..130G}. Origins of the seed particle populations invoked in the literature include the lower energy portions of material accelerated in energetic particle events associated with corotating interaction regions \citep[e.g.,][]{2017ApJ...838...23F}, CME-driven shocks \citep[e.g.,][]{2005ICRC....1..173L,2020ApJ...897..109Y}, solar flares \citep[e.g.,][]{2002ApJ...574.1039M,2014ASPC..484..234W}, and/or continuous acceleration processes occurring in interplanetary space \citep[see Section 5.3 in][ and references therein]{2016LRSP...13....3D}.

The suprathermal energy range is generally considered to extend from $\sim$10 keV/nuc to $\sim$1 MeV/nuc and to even higher energies \citep{2012SSRv..172..241M}. The particular energy range of the seed populations over which a shock acts may vary in different conditions or among different SEP events \citep{2019ApJ...872...89K}, and several authors have used the term ``suprathermal'' to refer to background populations extending up to $\sim$2 MeV/nuc. For example, \citet{2012AIPC.1500..128M} found that the maximum 12--80 MeV/nuc Fe fluences of SEP events at 1 au are apparently limited by the 0.04--1.81 MeV/nuc Fe number densities measured one day before the SEP events start. \citet{2014ApJ...784...47K,2014ApJ...791....4K} found a significant positive correlation between the $\sim$2 MeV proton intensities prior to the SEP event onsets at 1 au and the $\sim$20 MeV proton peak intensities of the events, which is consistent with these $\sim$2 MeV protons being part of the seed particle populations. However, considering the intensity of suprathermal particles near 1 au prior to an SEP onset does not improve the prediction of the following SEP peak intensity \citep{zhuang2021}. This may be due to the fact that the properties of suprathermal particles vary with the heliocentric distance, and the measurements near 1 au may differ significantly from those close to the Sun where particles are accelerated.

Observations close to the Sun can lead to a better understanding of suprathermal particles and their role as seed populations for SEP events \citep{2016SSRv..204..187M}. The NASA Parker Solar Probe (PSP) mission provides the first close ($<0.3$ au) direct exploration of our Sun and its environment \citep{2016SSRv..204....7F}. Since its launch in 2018, suprathermal and energetic particle events are available to be observed in the innermost heliosphere  \citep[e.g.,][]{2019Natur.576..223M,cohen2021,chhiber2021,2020ApJS..246...29G,2020ApJ...897..134L,2020ApJS..246...33S}. \citet{cohen2021} studied He/H abundances in a series of SEP events observed by PSP inside 0.5 au, and found dramatic event-to-event variations. \citet{2020ApJS..246...33S} analyzed the proton seed populations below 1 MeV when PSP was at $\sim$0.5 au. They found that the seed populations were enhanced during the passage of an interplanetary CME (ICME), which might indicate a key ingredient of the pre-acceleration process occurring close to the Sun. However, a number of still unresolved questions about the properties of seed particle populations close to the Sun include: (1) how long seed particle populations remain in the inner heliosphere before they are accelerated, (2) how widely seed particle populations are distributed in the heliosphere, and (3) what the differences are between these populations observed near 1 au and close to the Sun. 

To investigate the three questions above, we study the properties of $\sim$1--2 MeV protons associated with a series of SEP events observed by PSP in the inner heliosphere and the Solar TErrestrial RElations Observatory \citep[STEREO;][]{2008SSRv..136....5K} near 1 au, focusing on the characteristics of the decay phase of the particle enhancements at both locations. The decay phase of proton and electron intensities in large SEP events is of interest because they might indicate the formation of energetic particle reservoirs that fill broad regions of the heliosphere \citep[e.g.,][]{1972JGR....77.3957M,1992GeoRL..19.1243R}. \citet{1972JGR....77.3957M} found that the decay phase of SEP events can be separated into two clear phases based on the length of their decay time, i.e., the first characterized by a relatively short decay time and the second by a much longer decay time. Characteristics of the decay time of proton and electron intensities in SEP events as a function of solar cycle, heliolongitude, radial distance, solar wind speed, energy spectra, and interplanetary shock contributions have been previously analyzed by several authors \citep[e.g.,][]{2003AdSpR..32.2655D,2009JGRA..114.6102K,2010ApJS..189..181L}. 

The paper is organized as follows. The instrument and data used are introduced in Section \ref{instr}. The observation and analysis of a series of SEP events from 2020 May 27 to June 2 observed by PSP and STEREO are presented in Sections \ref{ev_psp} to \ref{1aucmp}. In this paper we follow previous studies by, e.g., \citet{2012AIPC.1500..128M}, \citet{2014ApJ...784...47K,2014ApJ...791....4K}, and \citet{2019ApJ...872...89K} to use $\sim$1--2 MeV as representative of the higher bound of the suprathermal energy range. Discussions and conclusions are given in Sections \ref{dis} and \ref{sum}, respectively.

\section{Successive CME and SEP Events}\label{data_ana}
\subsection{Instrument and Data}\label{instr}
The Integrated Science Investigation of the Sun (IS$\odot$IS) instrument suite \citep{2016SSRv..204..187M} on board PSP provides comprehensive measurements of SEPs. Here we use data from the IS$\odot$IS high-energy Energetic Particle Instrument (EPI-Hi). EPI-Hi is made of two low-energy telescopes (LET1 and LET2) and one high-energy telescope (HET), and measures particles in the energy range of $\sim$1--200 MeV/nuc \citep{2016SSRv..204..187M,2017ICRC...35...16W}. Both LET1 and HET are double-ended telescopes. During PSP's solar encounters, one LET1 aperture (LET-A) is nominally pointed $45^\circ$ west of the Sun-spacecraft line (i.e., along the nominal Parker spiral direction at 1 au), while its opposite end (LET-B) is nominally pointed $135^\circ$ east of the Sun-spacecraft line. As for HET, the two ends are nominally pointed $20^\circ$ west and $160^\circ$ east of the Sun-spacecraft line. We also use observations from the Low Energy Telescope \citep[LET,][]{2008SSRv..136..285M} on board STEREO-A (STA hereafter). 

To identify the solar eruptions associated with the origin of the SEP events, we use multiple instruments including the Large Angle and Spectrometric Coronagraph on board the SOlar and Heliospheric Observatory \citep[SOHO/LASCO;][]{1995SoPh..162..357B}, the Extreme Ultraviolet Imager \citep[EUVI;][]{2004SPIE.5171..111W} and the coronagraphs COR1 and COR2 \citep{2008SSRv..136...67H} on board STA. We use the Radio Frequency Spectrometer \citep[RFS;][]{2017JGRA..122.2836P} of the Electromagnetic Fields Investigation \citep[FIELDS;][]{2016SSRv..204...49B} on board PSP, and STA/WAVES \citep{2008SSRv..136..487B} for dynamic radio observations. CME propagation parameters are obtained from the Coordinated Data Analysis Workshop (CDAW) CME catalog based on LASCO observations \citep[\url{cdaw.gsfc.nasa.gov/CME_list/index.html};][]{2004JGRA..109.7105Y,2009EM&P..104..295G}. In addition, we use the Plasma and Suprathermal Ion Composition \citep[PLASTIC;][]{2008SSRv..136..437G} investigation on board STA and the Solar Wind Electrons Alphas and Protons \citep[SWEAP;][]{2016SSRv..204..131K} instrument suite on board PSP to estimate local solar wind speed. 

\subsection{SEP Events Observed by PSP in the Inner Heliosphere}\label{ev_psp}
From 2020 May 27 to 2020 June 2, a series of CMEs erupted, and led to a series of SEP events observed by PSP/IS$\odot$IS. These events have been studied by \citet{cohen2021} who focused on the variations of He/H abundances, and \citet{chhiber2021} who analyzed the role of magnetic field line random walk in the path length of SEPs. Figure \ref{psp_sta_source}(a) shows the locations of the Sun (orange), Earth (green) and STA (red), and the orbit of PSP (blue) from 2020 May 27 to 2020 June 3 in Heliocentric Earth Ecliptic (HEE) coordinates. The gray curves indicate the nominal Parker spiral magnetic field lines calculated using a constant solar wind speed of 320 km s$^{-1}$ as measured by STA/PLASTIC. During this period, PSP was making its fifth orbit about the Sun. Its distance from the Sun varied from 0.39 to 0.20 au, as shown in Figure \ref{psp_sta_source}(b). Meanwhile, STA was located at 0.97 au from the Sun and $72^\circ$ east from Earth, whereas PSP longitude ranged from $144^\circ$ to $182^\circ$ west of Earth.

Figure \ref{psp_isois}(b) shows the proton intensity observed by HET-A, HET-B, LET-A and LET-B from 2020 May 27 to 2020 June 3 in different energy ranges. The gray arrows indicate the onset of the 1--2 MeV proton intensity enhancements for the four observed SEP events identified using hourly resolution data \citep[more details can be found in][]{cohen2021}. The differences in the intensities observed by each side of the double-ended telescopes give us an estimation of the proton anisotropy during these events. Anisotropies were significant throughout the fourth event with onset on June 1, whereas the first three events with onsets on May 27, 28 and 29 were more isotropic especially during their decay phase. Figures \ref{psp_isois}(c) and (d) show the pitch angle scanned by the central axis of the apertures A (red) and B (blue) of the HET and LET1 telescopes, respectively. The variations in the pitch-angle scanned by these telescopes resulted from both variations of the direction of the interplanetary magnetic field but also from rolls performed by the spacecraft throughout this period. We note that variations of the pitch angle scanned by each telescope did not lead to significant changes in the particle intensities for these four SEP events.

\subsection{Contemporaneous Remote-Sensing and In Situ Observations}\label{ev_remote}
\begin{figure}[!hbt]
	\centering
	\includegraphics[width=\textwidth]{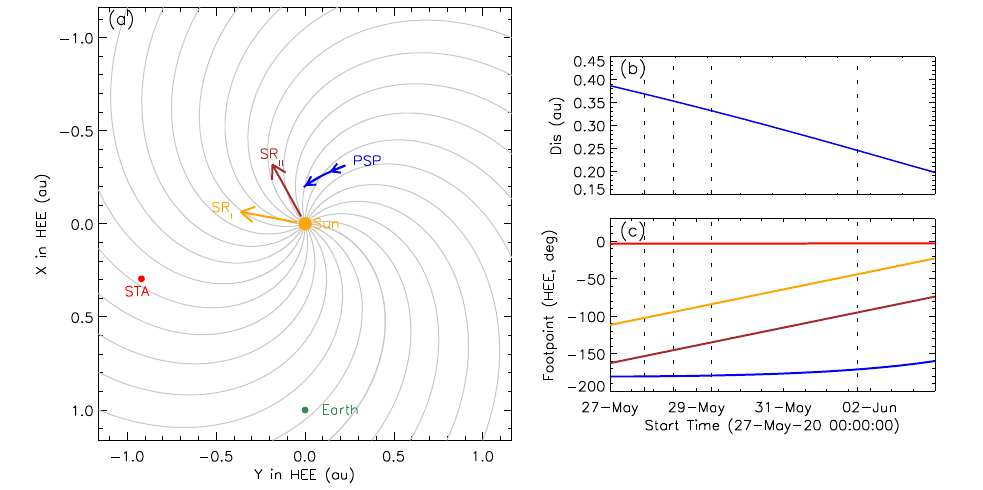}
	\caption{\small (a) Locations of the Sun (orange), Earth (green) and STA (red), and the orbit of PSP (blue) from 2020 May 27 to 2020 June 3 in HEE coordinates. The blue arrows indicate the orbit direction of PSP, while the orange and brown arrows show the longitude of the two SEP source regions on May 28. The gray curves show the nominal Parker spiral magnetic field lines. (b) Heliocentric distance of PSP. (c) Longitudes of source region I (orange), source region II (brown), and magnetic foot points of STA (red) and PSP (blue) on the solar surface. The dotted lines in (b) and (c) indicate the onset time of the four SEP events.} 
	\label{psp_sta_source}
\end{figure}

Based on the remote-sensing observations from SOHO/LASCO/C2, STA/EUVI in the 195\r{A} and 304\r{A} passbands, and STA/COR2, we identify eight major CME eruptions which may be related to the four SEP events. It is noted that only few CMEs can be well observed by LASCO/C2 and/or STA/COR2, but STA/EUVI observations show discernible eruptive signatures. These eruptions originated from two source regions: one in the northern hemisphere at N34, identified as $\rm{SR_{I}}$ in Figure \ref{event-image}(a), and the other in the southern hemisphere at S24, identified as $\rm{SR_{II}}$ in Figure \ref{event-image}(d). The vertical bars in Figure \ref{psp_isois}(a) indicate the onset times of these eruptions determined from STA/EUVI images after subtracting the $\sim$8.3 minutes of the light transit time from the Sun to STA (orange and brown lines indicate whether they were originated from $\rm{SR_{I}}$ or $\rm{SR_{II}}$, respectively). The longitude of the two source regions on May 28 are indicated by the arrows in Figure \ref{psp_sta_source}(a), and the variations of their HEE longitudes from 2020 May 27 to June 3 are shown in Figure \ref{psp_sta_source}(c). When the source regions co-rotated to the front side of the Sun, they were identified as active regions 12764 ($\rm{SR_{I}}$) and 12765 ($\rm{SR_{II}}$). Figure \ref{psp_sta_source}(c) also shows the temporal variation of the longitude of the magnetic footpoint of PSP on the solar surface (blue). The footpoint is estimated using nominal Parker spiral with a constant solar wind speed of 250 km s$^{-1}$ based on PSP/SWEAP data. It is found that the nominal PSP footpoint was quite distant from $\rm{SR_{I}}$, with longitudinal separation varying between $\sim$$69^\circ$ to $\sim$$137^\circ$. On May 27 the PSP footpoint was separated just by $18^\circ$ in longitude from $\rm{SR_{II}}$, but this longitudinal separation increased with time as the source region rotated away from PSP. Below we use the CME propagation parameters obtained from the CDAW catalog.

Based on hourly resolution data, the 1--2 MeV proton intensity onset and peak times during the first SEP event were observed at $\sim$19:00 UT and $\sim$23:00 UT on May 27. Figures \ref{event-image}(a), (b) and (c) show the CME (CME1) associated with the origin of the first SEP event and that was generated from $\rm{SR_{I}}$ as seen in STA/EUVI-304\r{A}, STA/COR2 and LASCO/C2 at three different times, respectively. According to STA/EUVI observations, the CME eruption occurred at 18:10 UT. Figure \ref{event-image}(a) also shows the filament associated with brightening footpoint on the solar surface. Type III radio bursts were observed by STA/WAVES in Figure \ref{event-image}(g) and PSP/FIELDS/RFS in Figure \ref{event-image}(h) at around 18:10 UT on May 27. This CME had a speed of 726 km s$^{-1}$ and an angular width of $51^\circ$. Note that an interplanetary intervening structure (i.e., ICME) was observed in situ by PSP during the period of this first SEP event, and this structure may have influenced the transport of SEPs to PSP (see Appendix \ref{icme}).

The onset and peak times of the 1--2 MeV proton intensities of the second SEP event occurred at $\sim$11:00 UT and $\sim$18:00 UT on May 28. We find three CMEs prior to this SEP onset, including two from $\rm{SR_{II}}$ (CME2 and CME3) and one from $\rm{SR_{I}}$ (CME4). The type III radio bursts associated with CME3 and CME4 were observed by both STA and PSP, while the one associated with CME2 was only visible in PSP/FIELDS/RFS (see Figures \ref{event-image}(g) and (h)). Figures \ref{event-image}(d) and (e) show the CME2 eruption in STA/EUVI-304\r{A} and STA/COR2, respectively. We note that the westward flux-rope like structure in Figures \ref{event-image}(e) and (f) is related to a different eruption from a source region located in the western and back hemisphere. We identify that the most likely contributor to the second SEP event was CME4, with an eruption time at 10:20 UT, a speed of 555 km s$^{-1}$ and an angular width of $44^\circ$. Most likely CME2 and CME3 did not contribute to the observed proton intensity enhancement, but they may have influenced the interplanetary magnetic fields along which the particles in this SEP event propagated.

The onset and peak times of the 1--2 MeV proton intensities of the third event occurred at $\sim$08:00 UT and $\sim$15:00 UT on May 29. Three successive CMEs from $\rm{SR_{I}}$ (CME5 to CME7) with associated type III radio bursts occurred within a short time interval on May 29 (Figures \ref{event-image}(g) and (h)). These successive eruptions may have contributed to a relatively longer decay time of this event as discussed in Section \ref{sec_decay}. As for CME5 to CME7, the eruption times were 06:46 UT, 10:46 UT and 16:46 UT, the CME speeds were 337, 347 and 210 km s$^{-1}$, and their angular widths were $37^\circ$, $52^\circ$ and $34^\circ$, respectively. Finally, the fourth SEP event, with onset time at $\sim$17:00 UT and peak time at $\sim$19:00 UT on June 1 for the 1--2 MeV proton intensities, is found to be associated with a CME from  $\rm{SR_{II}}$ (CME8). This CME had an eruption time at 16:26 UT, a speed of 248 km s$^{-1}$ and an angular width of $65^\circ$. The associated type III radio bursts can be found in \citet{cohen2021}.

Overall, the two source regions of the four SEP events were distant from the PSP nominal magnetic footpoint at the onset time of the SEP events, as indicated by the four vertical dotted lines in Figure \ref{psp_sta_source}(c). The parent CMEs were slow and quite narrow (in fact, CMEs 1 to 7 appeared like jets rather than flux ropes as also mentioned by \citet{cohen2021}). Table \ref{psp_sep} lists the information of these four SEP events associated with the CMEs and type III radio bursts.

\subsection{Decay Phase of SEP Events Observed by PSP}\label{sec_decay}
We analyze the decay phase of the four SEP events in the 1--2 MeV proton energy range as measured by LET-A. During each SEP event, the fast decay phase observed just after the SEP peak is neglected, and only the late decay phase is used. The shaded regions in Figure \ref{decay_time} show the periods selected to analyze their decay phases. We use an exponential function $J \propto \exp(-t/\tau_D)$ to fit the time-intensity profile, where $J$ is the proton intensity corrected for the background intensity ($J_{BKG}$) and pre-event intensity ($J_{PRE}$), and $\tau_D$ is the decay-time constant. In this paper, $J_{BKG}$ is calculated by averaging the 1--2 MeV proton intensity during a quiet period prior to the onset of the first SEP event. The result of $J_{BKG}$ is $\sim$$2 \times 10^{-3}$ $\rm{proton} \ \rm{cm}^{-2} \ \rm{s}^{-1} \ \rm{sr}^{-1} \ \rm{MeV}^{-1}$, as indicated by the horizontal black line in Figure \ref{decay_time}. The pre-event intensity correction is significant for the second to the fourth events. $J_{PRE}$ at a specific time $t_0$ for an event (event number $i=2$, 3 or 4) is estimated by extending the exponential fit result(s) from preceding event(s) to $t_0$, i.e., $J_{PRE}(t_0)=\sum_{n=1}^{i-1}J_{0n} \exp{(-t_0/\tau_{Dn})}$, where $J_{0n}$ and $\tau_{Dn}$ are the fitted results for the nth preceding event. The orange profile refers to the original data points, and the four purple profiles in the shaded regions indicate the corrected $J$. We note that for the third SEP event, there may exist small additional enhancements by successive eruptions of CME6 and CME7. However, the disturbed profile may bring uncertainties in determining the late phase of each small enhancement, and the short time interval between successive CMEs (5--7) indicates that a series of injections may be (partly) merged. Therefore, the data points in the third shaded region are used as a whole for measuring the related decay-time constant.

The results of the exponential fit are shown by the straight black lines in Figure \ref{decay_time}. $\tau_D$ is fitted to be 15 hours for the first, 19 hours for the second, 22 hours for the third, and 7 hours for the fourth SEP event, which are also listed in Table \ref{psp_sep}. Starting from a transport model by considering the dominant roles of convection and adiabatic deceleration in the solar wind, and neglecting the terms of gradients, scattering and drift, $\tau_D$ would be estimated by
\begin{equation}\label{eq1}
	\tau_D=\frac{3}{2+\alpha \gamma} \frac{R}{2V_{SW}},
\end{equation}
where $R$ is the heliocentric distance, $V_{SW}$ is the solar wind speed, $\gamma$ is the differential exponent of the energy spectrum, and $\alpha \approx 2$ at nonrelativistic energies \citep[see][]{1970JGR....75.3147F,2003AdSpR..32.2655D,2009JGRA..114.6102K}. The limitations of using Equation \ref{eq1} were discussed by \citet{2009JGRA..114.6102K}. We focus here on the decay phase of the first three events which were from the same source region, and the corresponding $R$, $V_{SW}$ and $\gamma$ (using differential proton intensity in 1--2 MeV with correction for the background and pre-event) are measured at the initial time of each decay phase. For the first three events, $R$ was 0.363, 0.344 and 0.321 au, $V_{SW}$ was 280, 290 and 280 km s$^{-1}$, and $\gamma$ was 2.5, 2.5 and 2.1, respectively. Therefore, $\tau_D$ derived by Equation \ref{eq1} for the first to the third decay phases is 12, 11 and 12 hours. For the third SEP event, $\tau_D$ estimated by the exponential fit is larger than those for the first two events, and larger than the value derived by Equation \ref{eq1}. Whereas a more diffusive energetic particle transport may contribute to a longer decay phase, it is also possible that a prolonged injection or even discrete injections by successive CMEs occurring within a short time interval contributed to the longer decay in this third SEP event (see discussion of the factors that play a role in the decay phase of the SEP events in Section 2 of \citet{2010ApJS..189..181L} and references therein). Ion intensities measured at PSP increased after the occurrence of CME5 (Figure \ref{psp_isois}), but we cannot discard that CMEs 6 and 7 contributed into extending the decay phase of the third event, although their occurrence did not translate into obvious individual particle intensity enhancements in an already elevated particle intensities at PSP.

\subsection{SEP Events Observed by STA near 1 au}\label{1aucmp}
\begin{figure}[!hbt]
	\centering
	\includegraphics[width=0.75\textwidth]{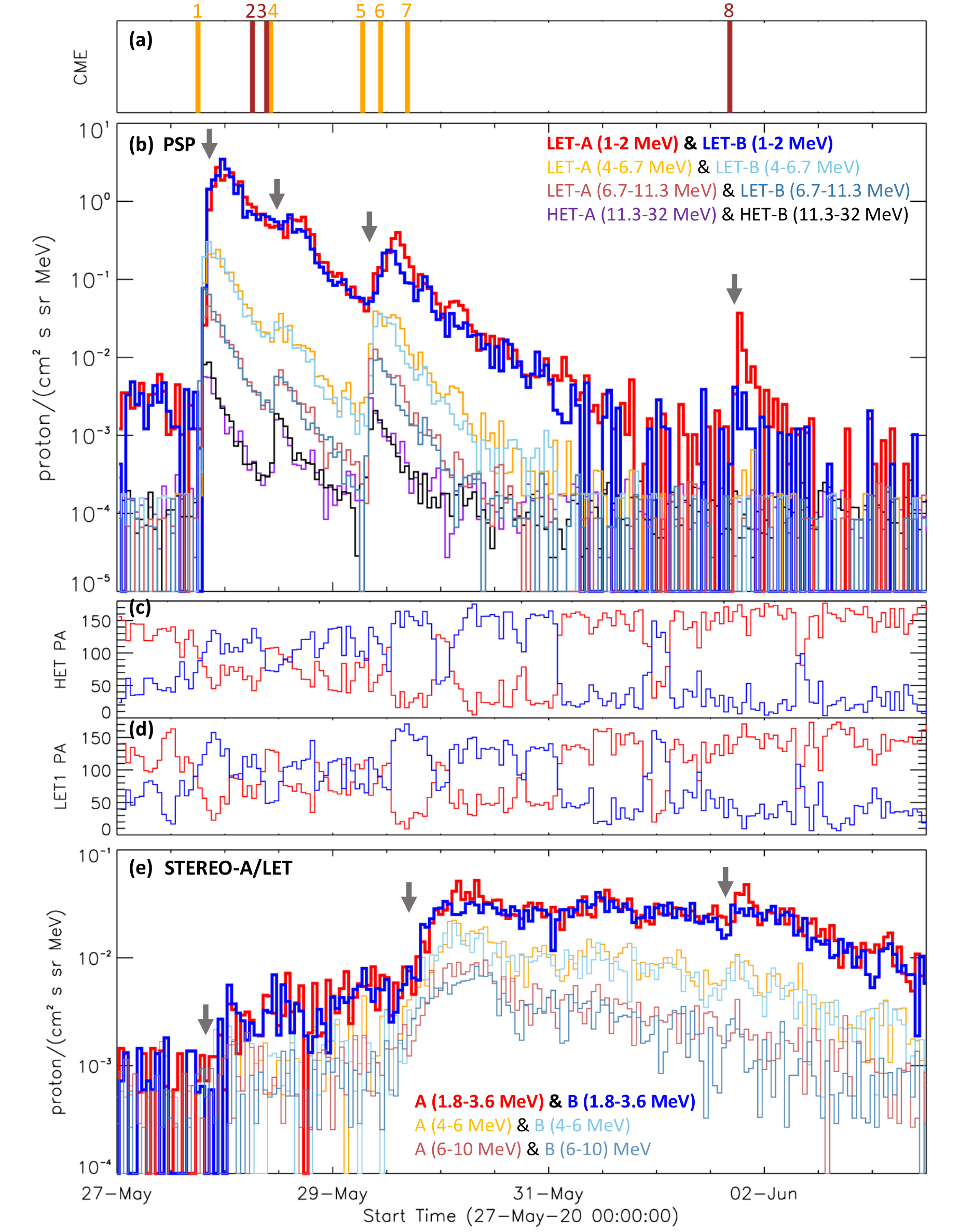}
	\caption{\small (a) Eruption time of the CMEs (with 8.3 minutes of the light transit time from the Sun to 1 au subtracted) from source region I (orange) and source region II (brown). (b) Temporal variation of the hourly proton intensity observed by HET-A, HET-B, LET-A and LET-B of PSP/IS$\odot$IS/EPI-Hi. (c)--(d) Pitch angle scanned by the axis of symmetry of the apertures A (red) and B (blue) of the telescopes (c) HET and (d) LET1. (e) Temporal variation of the hourly proton intensity observed by STA/LET-A and STA/LET-B. The arrows in (b) mark the 1--2 MeV proton onset time of the four SEP events observed by PSP. The arrows in (e) mark the 1.8--3.6 MeV proton onset time of the first, third and fourth SEP events observed by STA.}
	\label{psp_isois}
\end{figure}

\begin{figure}[!hbt]
	\centering
	\includegraphics[width=0.85\textwidth]{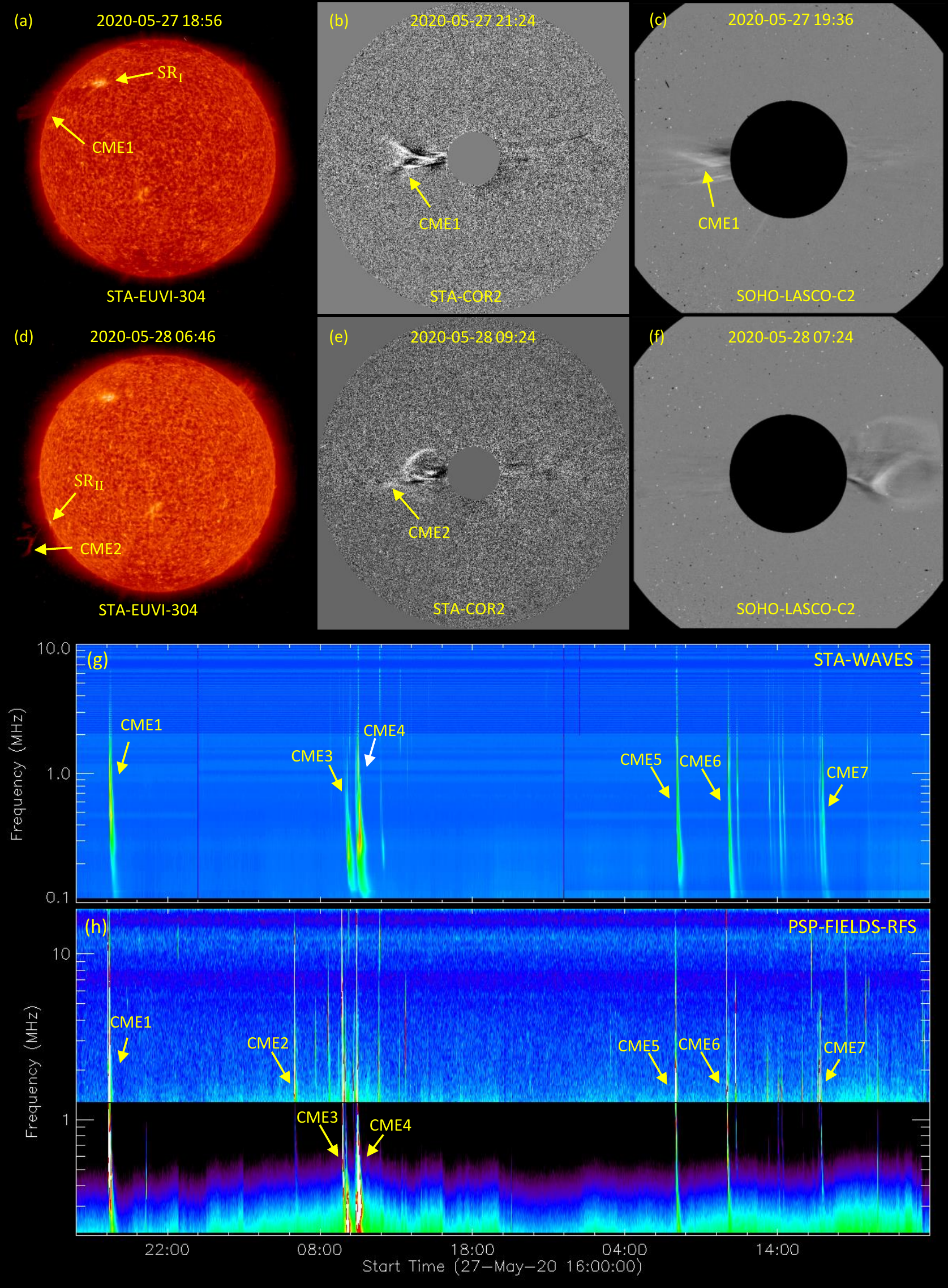}
	\caption{\small (a) and (d) STA/EUVI-304\r{A} observations. (b) and (e) Running-difference images observed by STA/COR2. (c) and (f) Running-difference images observed by SOHO/LASCO/C2. (g) and (h) Dynamic radio spectra observed by STA/WAVES and PSP/FIELDS/RFS.}
	\label{event-image}
\end{figure}

\begin{figure}[!hbt]
	\centering
	\includegraphics[width=\textwidth]{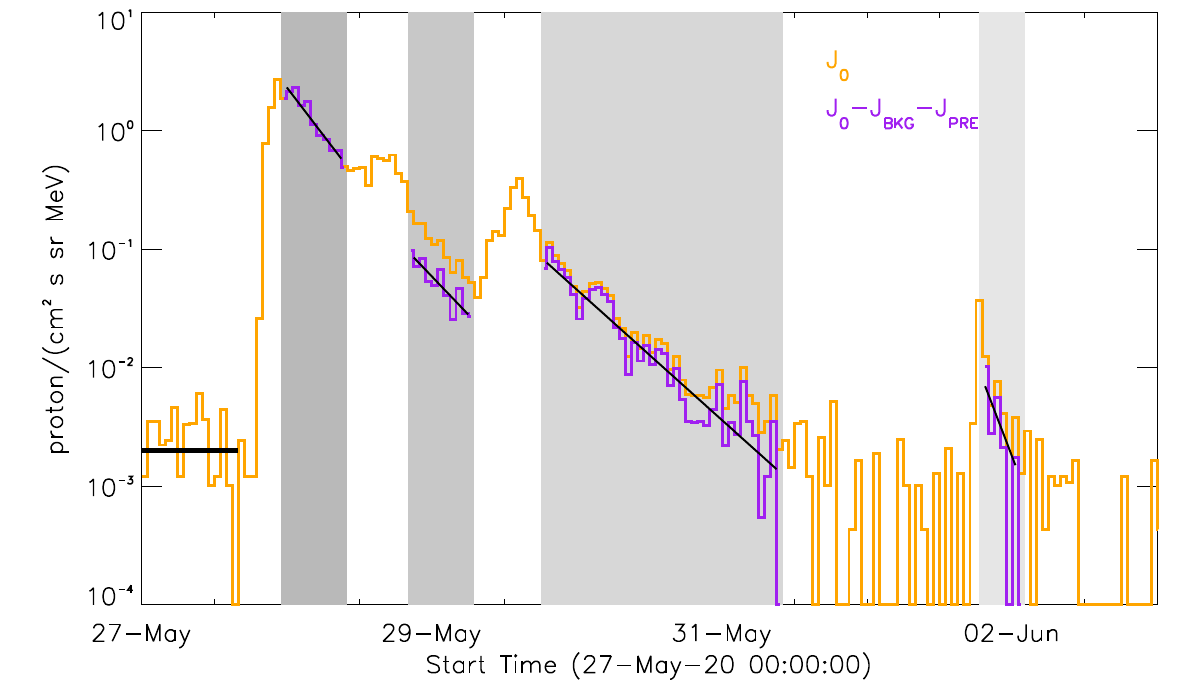}
	\caption{\small Decay phase of the four SEP events in 1--2 MeV energy range observed by PSP/IS$\odot$IS/LET-A. The orange profile refers to original proton intensity, and the purple profile is the background-corrected and pre-event-corrected intensity. The shaded regions indicate the selected late decay phase during the four SEP events.}
	\label{decay_time}
\end{figure}

STA/LET near 1 au also measured the SEP enhancements. Figures \ref{psp_sta_source}(a) and \ref{psp_sta_source}(c) show the location of STA and the longitude of its nominal magnetic footpoint by using a Parker spiral with a constant solar wind speed of 320 km s$^{-1}$. In Appendix \ref{sta_obs} we describe the particle intensity enhancements measured by STA and the contextual magnetic field and solar wind observations at this spacecraft. Figure \ref{psp_isois}(e) shows the hourly proton intensity from May 27 to June 3 in three different energy ranges observed by LET-A and LET-B, respectively. During this period, the center axis of LET-A was pointed $\sim$$45^\circ$ east of the Sun-spacecraft line, and the center axis of LET-B was pointed $\sim$$135^\circ$ west of the Sun-spacecraft line. The arrows in Figure \ref{psp_isois}(e) identify the onset of the 1.8--3.6 MeV proton intensity enhancements that could be associated with the first, third, and fourth SEP events observed by PSP. The absence of major eruptions near the central meridian (i.e., from solar regions magnetically well connected to STA) during this time as seen from Solar Dynamic Observatory/Atmospheric Imaging Assembly observations, and the coincidence of the particle intensity enhancement onsets at STA with the solar eruptions described in Table \ref{psp_sep}, suggest a common origin for the events at STA and at PSP. However, the presence of solar wind structures observed in situ by STA may have influenced the time intensities observed by STA as described in Appendix B.

In general, Figure \ref{psp_isois} shows that the proton intensity enhancements observed by STA/LET were lower than those at PSP, especially for the first two events by about two to three orders of magnitude (although intercalibration among different instruments is needed). With the exception of the peak intensities of the third and fourth events at STA, anisotropies at STA were quite low. During this time, STA was located to the west of both $\rm{SR_{I}}$ and $\rm{SR_{II}}$, and the angular separations between its nominal magnetic footpoint and the two source regions became smaller due to the rotation of the Sun (cf. Figure \ref{psp_sta_source}(c)).

The third SEP event at STA displayed the largest proton intensities in the three energy ranges of LET. The onset of the event at 1.8--3.6 MeV proton intensities occurred around at 14:00 UT on May 29, that is six hours later than the onset time observed by PSP. The larger intensity enhancement in this event at STA with respect to the other two prior SEP events (all originated from $\rm{SR_{I}}$) may be due to the fact that the STA magnetic footpoint approached $\rm{SR_{I}}$. In Figure \ref{psp_isois}(e) the intensity profiles of this SEP event decline very slowly and extend to June 1 until the onset of the fourth SEP event. Discrete enhancements corresponding to the multiple injections by the three successive CMEs in the third event were difficult to distinguish at STA. \citet{2006ApJ...650.1199W} found that successive and discrete electron enhancements observed by Helios spacecraft at $\sim$0.3 au (though more impulsive with decay time of few hours) may become merged and unresolved by near-Earth spacecraft, which may be due to interplanetary particle transport processes. Similarly, using the exponential fit as done in Section \ref{sec_decay}, the decay-time constant of the third SEP event in 1.8--3.6 MeV by STA/LET-A is estimated to be $\tau_D > 200$ hours (only using the decay phase before the onset of the proton intensity enhancement on June 1), which is around ten times larger than that measured at PSP. In a higher energy range of 6--10 MeV at STA vs. 6.7--11.3 MeV at PSP, the $\tau_D$ estimate shows a similar result, i.e., 145 hours vs. 23 hours. The consideration of Equation \ref{eq1} is not enough to explain the difference of $\tau_D$, because the major contribution $R$ in Equation \ref{eq1} would only lead to an increase of $\tau_D$ by a factor of $\sim$3 from PSP to STA. We provide a possible explanation for such a difference in the next section.

\section{Discussion}\label{dis}
In general, the properties of the intensity-time profiles of SEP events depend on a mixture of two major factors, i.e., particle injection and transport. Injection may involve prompt particle injection by a flare and/or continuous injection by a CME-driven shock. If an impulsive injection occurs in a localize source (e.g., the flare site), the decay phase may exhibit a very fast decline; if the injection is extended over a broad longitudinal range, the decay may last longer. During the decay phase of an SEP event, the observer does not stay in the same magnetic flux tube along which particles propagate. Thus, the decay phase also depends on the state of interplanetary plasma and magnetic field, and on the location of the observer relative to the SEP source region. Furthermore, the effects of adiabatic deceleration and solar wind convection, which are a function of the SEP power-law energy spectrum and the solar wind speed, shall be considered (Equation 1). However, the eruptions at the origin of the SEP events in this paper occurred from regions distant from the nominal magnetic footpoints of both spacecraft. The associated CMEs were relatively small. This suggests that cross-filed transport processes may have played a role in spreading the particles during these events. The contribution of cross-field transport processes to the decay phase of the events is difficult to measure without a proper knowledge of the properties of the interplanetary medium and the characteristics of the particle injection. Besides, for the third SEP event the decay-time constant measured by STA is found to be more than one order of magnitude larger than that measured by PSP, which can not be simply explained by Equation \ref{eq1}.

\begin{figure}[!hbt]
	\centering
	\includegraphics[width=\textwidth]{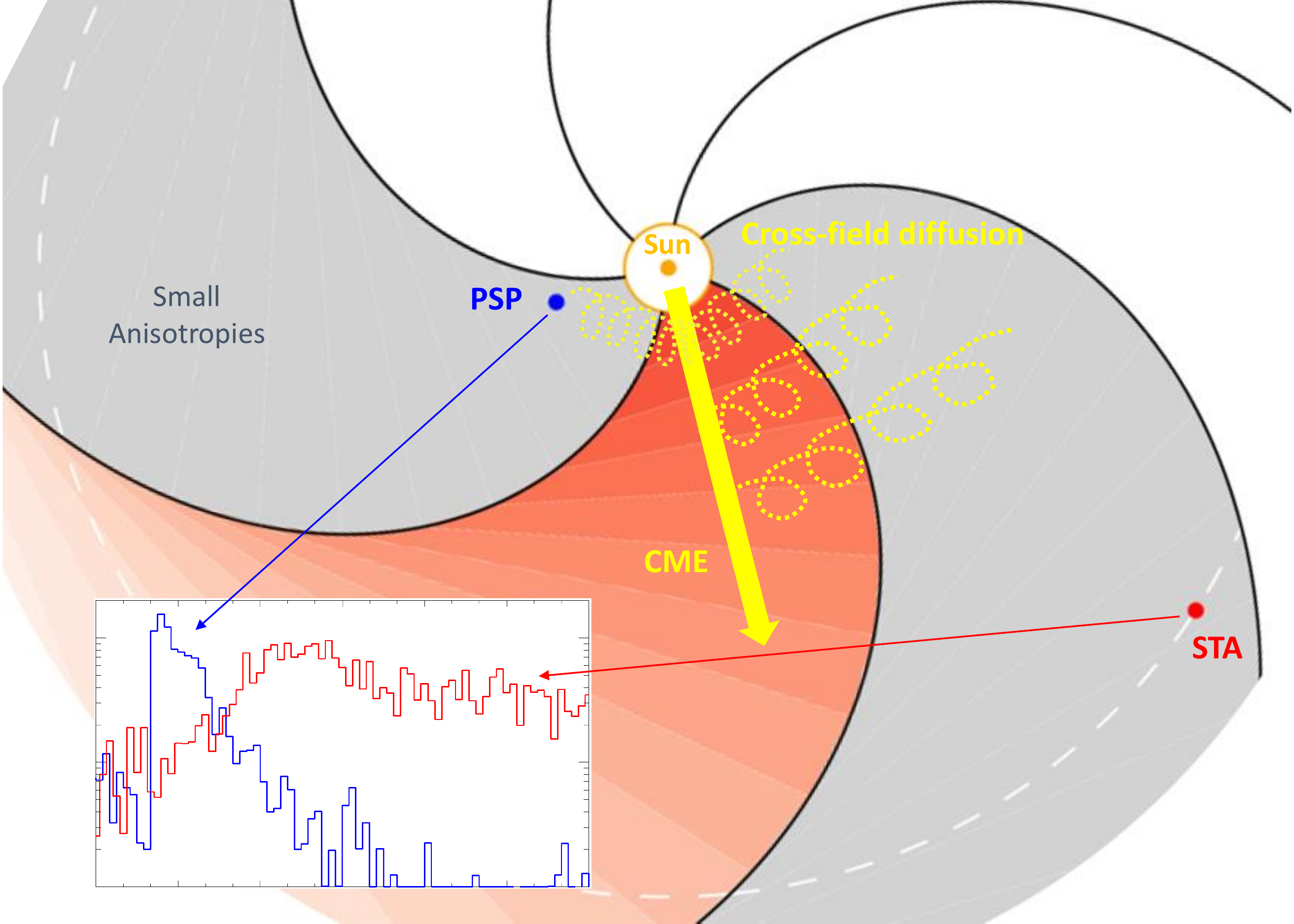}
	\caption{\small A cartoon showing different observations of suprathermal particles between PSP and STA. The reddish area marks the region with large anisotropies, and the gray areas are the regions where small anisotropies are measured. The yellow thick arrow indicates the propagation direction of the nose of the parent CME, and the yellow dotted helices refer to particle transport due to cross-field diffusion processes. The insert shows the proton intensities of the third SEP event observed by PSP/LET-A (6.7--11.3 MeV) and STA/LET-A (6--10 MeV). The figure is adapted from \citet{2014A&A...567A..27D}.}
	\label{scheme}
\end{figure}

Figure \ref{scheme} (adapted from \citet{2014A&A...567A..27D}) illustrates a possible scenario for this third event. The thick yellow arrow indicates the propagation direction of the nose of the parent CME; the reddish area marks the region where large anisotropies are usually observed, and the gray areas are the regions where small anisotropies are usually measured (as per \citet{2014A&A...567A..27D} discussion). PSP (blue) and STA (red) are located at different helioradii and both in the regions associated with small anisotropies. In the absence of broad particle sources, the transport of particles from the acceleration region to the magnetic field lines connecting both spacecraft is likely due to cross-field diffusion processes (indicated by the yellow dotted helices) because the nominal magnetic footpoints of these field lines were distant from the sites of the solar eruptions identified in Section \ref{ev_remote}. If a CME is present, as it propagates in interplanetary space, it may continue accelerating and injecting particles, acting like a continuous source. Since STA is located near 1 au compared to PSP located in the inner heliosphere, it may be able to observe a more prolonged injection of particles as long as the shock driven by the CME continues to efficiently accelerate particles as it approaches the observer. Furthermore, since PSP and STA were on the east and west of the SEP source region separately, the rotation of the Sun may also be considered as the magnetic flux tubes filled with accelerated particles move toward STA, implying a more extended SEP event at STA than at PSP with a longer decay time. We have symbolized these processes in Figure 5 by the different number of helices toward PSP and STA.

This work shows that slow and narrow CMEs, e.g., with speed around 300 km s$^{-1}$ and angular width $<60^\circ$, can result in the acceleration of particles to at least $\sim$1 MeV and even above 10 MeV as in the first three SEP events. Additional cases were found by \citet{2020ApJS..246...33S}, in which the enhancements of suprathermal/energetic protons associated with relatively slow and narrow CMEs were observed by PSP at $\sim$0.5 au. As for particle acceleration related to slow and narrow CMEs, the absence of type II radio bursts shown in Figures \ref{event-image}(g) and (h) indicates that the CME-driven shocks were not efficient enough to produce electrons responsible for the radio emission but able to efficiently accelerate $\sim$10 MeV protons. Note, however, that $>$0.5 MeV electrons were observed by PSP in the first and third SEP events \citep{cohen2021}, suggesting that the acceleration of particles in these events may be related to CME-associated magnetic reconnection or compression processes. In our SEP events, simultaneous observations by PSP and STA with separation of $\sim$$140^\circ$ in longitude reveal that the accelerated particles may fill up a large region in longitude in the inner heliosphere; the analysis of the event decay phase indicates that these particles can remain there for around one day. These widespread and one-day lasting particles generated by successive eruptions associated with slow and narrow CMEs in the inner heliosphere may act as ubiquitous seed populations in large SEP events. However, since the enhancement level of proton intensities of the SEP events at STA is found to be much lower than that at PSP, it indicates that there may still exist a large population of suprathermal seed particles in the inner heliosphere that we cannot observe with current instrumentation near 1 au, which limits the ability of forecasting large SEP events by using suprathermal particle intensity measured near 1 au.

\section{Summary}\label{sum}
We studied four successive SEP events with proton intensity enhancements observed in the $>$1 MeV energy range from 2020 May 27 to 2020 June 2 by PSP at distances between $\sim$0.4 and $\sim$0.2 au and STA near 1 au. We identified the CME eruptions associated with these SEP events. We found those SEP events were widespread in the heliosphere, associated with small anisotropies of proton intensities observed by both PSP and STA. We investigated the decay phase of each SEP event at PSP in the $\sim$1--2 MeV proton energy range, and their decay-time constants were found to be around ten hours to one day, which may indicate the properties of suprathermal seed particles in the inner heliosphere. For the third event on May 29, we found that the longer decay time may be caused by continuous particle injections associated with three successive CMEs within a short time interval. We compared the observational characteristics (enhancement magnitude and decay phase) of proton intensities measured by PSP and STA at different heliocentric distances. We provided a probable illustration to explain that during the third SEP event the decay time at STA was much longer than that at PSP. These results highlight the importance of measuring and understanding seed particles in the inner heliosphere, as the intensities measured in the inner heliosphere by PSP differed significantly from those measured near 1 au by STA.

\textbf{Acknowledgement} Parker Solar Probe was designed, built, and is now operated by the Johns Hopkins Applied Physics Laboratory as part of NASA’s Living with a Star (LWS) program. Support from the LWS management and technical team has played a critical role in the success of the Parker Solar Probe mission. We acknowledge the IS$\odot$IS, SWEAP and FIELDS Science Teams. We also thank the use of the data from SOHO and STEREO. The data used in this paper can be downloaded from \url{spdf.gsfc.nasa.gov} and \url{sdac.virtualsolar.org/cgi/search}. The CME catalog used in this paper are generated and maintained at the CDAW Data Center by NASA and the Catholic University of America in cooperation with the Naval Research Laboratory. Research for this work was made possible by NASA grants 80NSSC20K0431, 80NSSC17K0009 and 80NSSC19K0831. D.L. acknowledges support from NASA Living With a Star (LWS) programs NNH17ZDA001N-LWS and NNH19ZDA001N-LWS.

\clearpage

\begin{sidewaystable}[!htb]
	\centering
	\caption{A series of SEP events observed by PSP/IS$\odot$IS associated with CMEs and type III radio bursts from 2020 May 27 to June 2. }
	\begin{tabular}{|c|c|c|c|c|c|c|c|c|c|c|}
		\hline
		\multicolumn{3}{|c|}{SEP} & Source & \multicolumn{5}{c|}{CME} & \multicolumn{2}{c|}{Type III} \\ \cline{1-3} \cline{5-11}
		Onset & Peak & $\tau_D$ & Region & \# & Onset & CPA & $W$ & $V$ & PSP & STA \\ \cline{1-11}
		2020/05/27 19:00 & 23:00 & 15 & $\rm{SR_{I}}$ & 1 &05/27 18:10 & $80^\circ$ & $51^\circ$ & 726 & 18:05 & 18:10 \\ \cline{1-11}
		2020/05/28 11:00 & 18:00 & 19 & \makecell{$\rm{SR_{II}}$ \\ $\rm{SR_{II}}$ \\ \textbf{$\rm{SR_{I}}$}} & \makecell{2 \\ 3 \\ \textbf{4}} & \makecell{05/28 06:18 \\ 05/28 09:26 \\ \textbf{05/28 10:20}} & \makecell{- \\ - \\ \textbf{76$^\circ$}} & \makecell{- \\ - \\ \textbf{44$^\circ$}} & \makecell{- \\ - \\ \textbf{555}} & \makecell{06:20 \\ 09:25 \\ \textbf{10:20}} & \makecell{- \\ 09:30 \\ \textbf{10:20}} \\ \cline{1-11}
		2020/05/29 08:00 & 15:00 & 22 & \makecell{$\rm{SR_{I}}$ \\ $\rm{SR_{I}}$ \\ $\rm{SR_{I}}$} & \makecell{5 \\ 6 \\ 7} & \makecell{05/29 06:46 \\ 05/29 10:46 \\ 05/29 16:46} & \makecell{$79^\circ$ \\ $84^\circ$ \\ $76^\circ$} & \makecell{$37^\circ$ \\ $52^\circ$ \\ $34^\circ$} & \makecell{337 \\ 347 \\ 210} & \makecell{07:20 \\ 10:40 \\ 16:45} & \makecell{07:20 \\ 10:45 \\ 16:45} \\ \cline{1-11}
		2020/06/01 17:00 & 19:00 & 7 & $\rm{SR_{II}}$ & 8 & 06/01 16:26 & $79^\circ$ & $65^\circ$ & 248 & 16:30 & 16:40 \\ \cline{1-11}
	\end{tabular}	
	\label{psp_sep}
	\footnotesize
	\begin{tablenotes}
		\item[1] [1] The table lists the SEP onset time, peak time and decay-time constant ($\tau_D$, in unit of hour) for proton intensity in the 1--2 MeV energy range, the source region, the CME number, the CME eruption time based on STA/EUVI observations, the CME 2-dimensional parameters (CPA is the central position angle, $W$ is the angular width, and $V$ is the speed in unit of km s$^{-1}$) based on the CDAW catalog, and the onset time of type III radio bursts observed by PSP and STA.
		\item[2] [2] For the second event, the CME eruption in bold font is supposed to be the main driver for this event.
	\end{tablenotes}
	\normalsize
\end{sidewaystable}

\clearpage

\appendix
\section{Arrival of an ICME at PSP}\label{icme}
\begin{figure}[!hbt]
	\centering
	\includegraphics[width=0.9\textwidth]{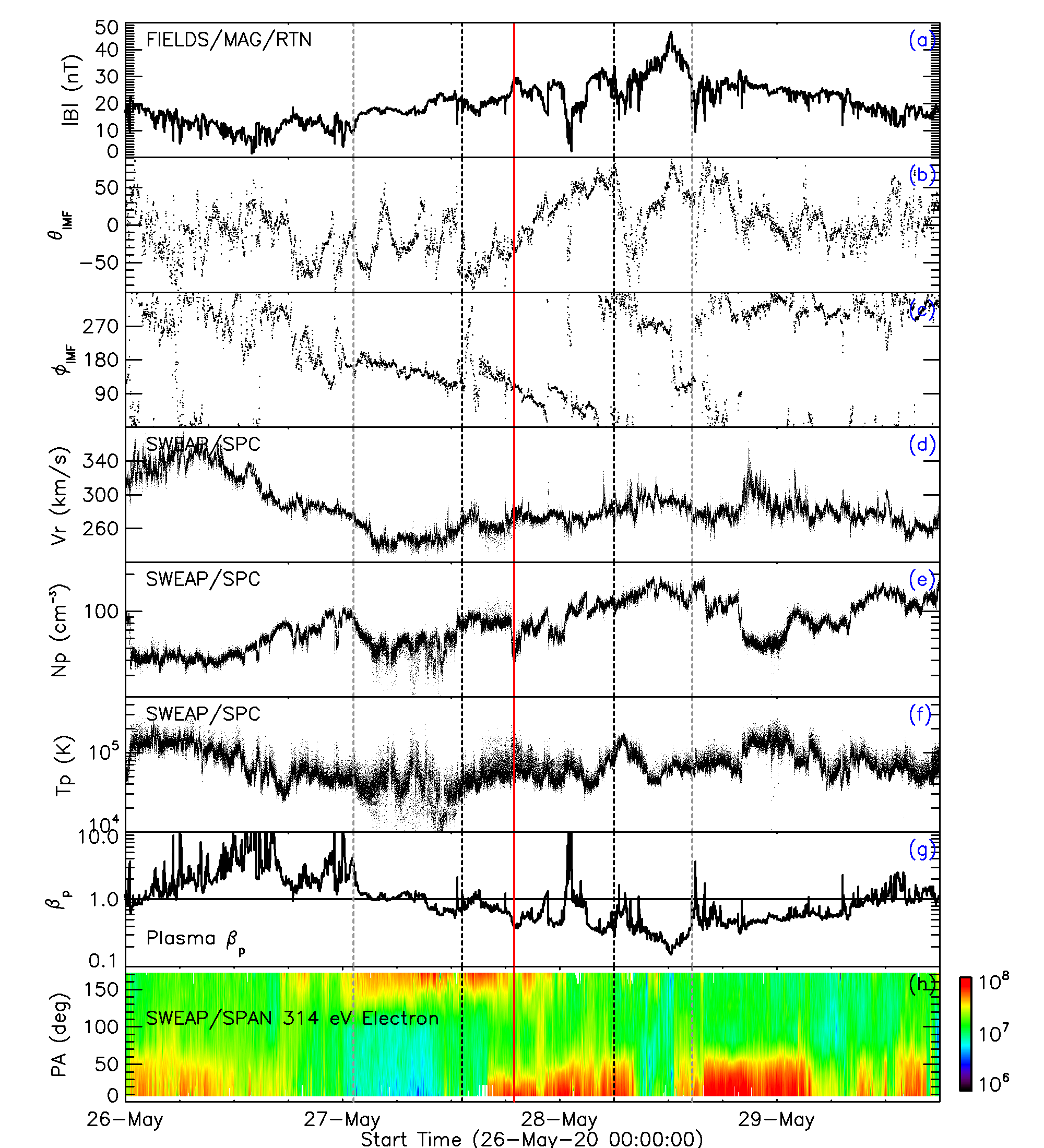}
	\caption{\small (a) Magnetic field strength. (b) Elevation angle of magnetic field in RTN coordinates. (c) Azimuth angle of magnetic field in RTN coordinates. (d) Proton radial velocity in RTN coordinates. (e) Proton number density. (f) Proton temperature. (g) Proton plasma beta $\beta_p$. (h) 314 eV electron pitch angle distribution. The two outer dashed gray lines mark the whole ICME region, while the inner dashed black lines mark the ICME core region. The vertical red line indicates the onset time of the first SEP event on May 27.}
	\label{psp_icme}
\end{figure}

From May 27 to 28, there was an ICME arriving at PSP. Figure \ref{psp_icme} shows the in situ observations of the (a) magnetic field strength, (b) elevation and (c) azimuth angles of magnetic field in the spacecraft centered Radial Tangential Normal (RTN) coordinate system, (d) proton radial velocity in RTN coordinates, (e) proton number density, (f) proton temperature, (g) proton plasma beta $\beta_p$, and (h) 314 eV suprathermal electron pitch angle distribution by FIELDS and SWEAP, respectively. 
The two outer dashed gray lines mark the region of the ICME between 01:12 UT on May 27 and 14:40 UT on May 28, while the two inner black lines indicate the core region of this ICME between 13:12 UT on May 27 and 06:00 UT on May 28. In the core region typical ICME features can be observed, including (1) the smooth rotation of magnetic field vectors, (2) the lower plasma beta, and (3) the bidirectional suprathermal electrons. Outside the core region, the elevation angle of the ICME was disturbed (but low $\beta_p$ was maintained), which may be due to the interaction of the ICME with the solar wind. Intervening ICMEs may affect the transport of SEPs by modifying interplanetary magnetic fields or providing a temporary magnetic connection between the SEP source region and the observer \citep[e.g.,][]{2012A&A...538A..32M,Palmerio_2021}. \citet{chhiber2021} used the velocity dispersion analysis (VDA) technique to estimate the path length of protons transporting from corona to PSP during the first SEP event, and found that this path length was around 0.625 au, significantly larger than the nominal Parker spiral length ($\sim$0.37 au) when PSP was at $\sim$0.35 au from the Sun. They explained the longer particle transport length by field line random walk or pitch angle scattering. However, the longer path length may also be due to the ICME traveling through PSP during the SEP event (the onset time of the first SEP event was roughly in the middle of the ICME intervening period, as indicated by the vertical red line in Figure \ref{icme}).

\section{STEREO-A Observations}\label{sta_obs}
\begin{figure}[!hbt]
	\centering
	\includegraphics[width=0.9\textwidth]{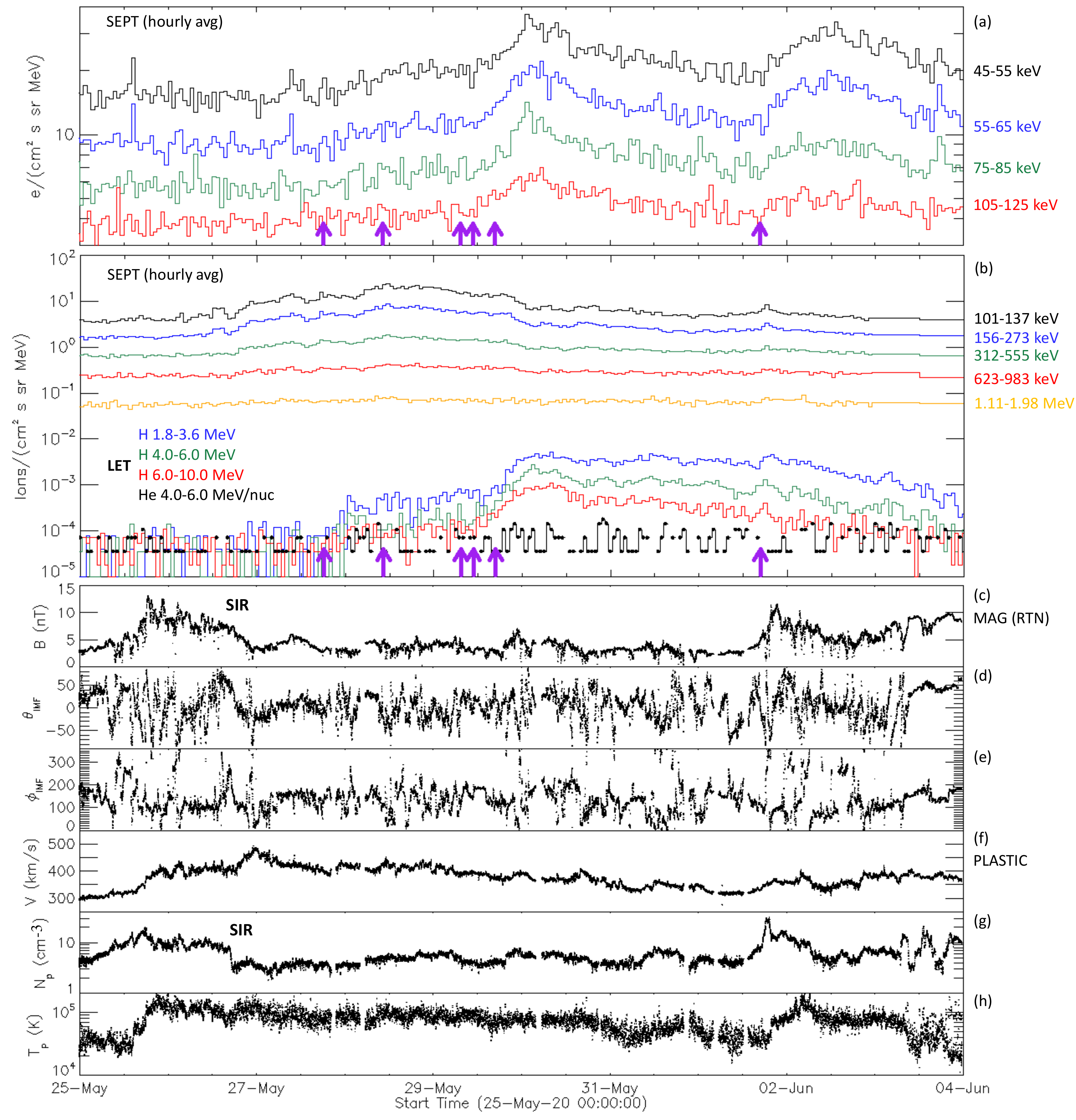}
	\caption{\small (a) Hourly electron intensity averaged over the four SEPT telescopes. (b) Hourly ion intensity averaged over the four SEPT telescopes (five top traces) and sector-averaged proton intensity measured by LET (blue, green, and red bottom traces) and sector-averaged 4-6 MeV/nuc He intensity measured by LET (discrete black symbols). (c) Magnetic field magnitude. (d) Elevation angle of magnetic field in RTN coordinates. (e) Azimuth angle of magnetic field in RTN coordinates. (f) Solar wind proton speed. (g) Proton number density. (h) Proton temperature. The purple arrows indicate the occurrence of the type III radio bursts seen by STA as listed in Table \ref{psp_sep}.}
	\label{sta_figure}
\end{figure}

In this appendix we describe in detail the STA measurements collected from May 25 to June 4. Figure \ref{sta_figure} shows from top to bottom (a) hourly averages of the electron intensities averaged over the four telescopes of the Solar Electron and Proton Telescope \citep[SEPT;][]{2008SSRv..136..363M}; (b) ion intensities averaged over the four SEPT telescopes (five top traces) and proton intensities averaged over all sectors measured by LET (blue, green, and red bottom traces) and 4-6 MeV/nuc He intensities averaged over all sectors measured by LET (discrete black symbols); magnetic field (c) magnitude, (d) elevation angle and (e) azimuth angle in RTN coordinates as measured by the Magnetic Field Experiment on board STA \citep[MAG;][]{2008SSRv..136..203A}; solar wind proton (f) speed, (g) number density, and (h) temperature as measured by STA/PLASTIC. The purple arrows indicate the occurrence of the type III radio bursts seen by STA as listed in Table \ref{psp_sep}.

Several particle intensity enhancements were observed during the time interval shown in Figure \ref{sta_figure}. A solar wind stream interaction region (SIR) crossed STA between 2020 May 25 and 26, as clearly seen by the compressed magnetic field and solar wind density and identified by the label SIR in Figures \ref{sta_figure}(c) and \ref{sta_figure}(g). This SIR was followed by a relatively high-speed ($\sim$500 km s$^{-1}$) solar wind stream on 2020 May 27. Low-energy ($\lesssim$600 keV) ion intensities showed a weak enhancement from the end of May 26 until May 30 throughout the passage of this relative fast solar wind stream as typically observed for particle events associated with SIRs \citep[e.g.,][]{1999SSRv...89...77M}. The proton intensities at energies between 1.8 MeV and 10 MeV observed by LET raised at the end of May 27 after the occurrence of CME1, whereas electrons at energies below $\sim$100 keV also showed a weak enhancement after this solar event. We note that SEPT has a relatively higher background level in the $\sim$1 MeV ion energy range so that the proton intensity enhancements seen by LET, if extending to low energies, cannot be resolved. It is also possible that part of the intensity enhancement seen by LET had contributions from the SIR event, although delayed with respect to the intensity enhancement at lower energies \citep[e.g.,][]{2021ApJ...908L..26W}. 

The occurrence of CMEs 2--4 on 2020 May 28 did not result in a new particle enhancement at STA, but the solar events on 2020 May 29 clearly caused a new particle enhancement observed by LET and at near-relativistic electrons. Similar to PSP observations, we cannot discern the individual contribution of CMEs 5--7 into this particle enhancement, but the lack of concomitant solar eruptions from regions where STA would establish nominal magnetic connection and the time coincidence seem to indicate a common origin for the third SEP event at PSP and the main event at STA. The discrete 4--6 MeV/nuc He intensities also showed an increase associated with this SEP event. Estimation of the He/H abundance ratio measured during this intensity enhancement, from 16:00 UT on May 29 to 00:00 UT on June 1, provides a value of $\sim$0.015-0.05 over the energy range 4-8 MeV/nuc that is found to be consistent with the He/H ratio of the SEP event on May 29 observed by PSP (0.016) as described in \citet{cohen2021}.

Finally, the occurrence of CME-8 on 2020 June 1 was accompanied by an enhancement of near-relativistic electron intensities and a weak enhancement superposed on the already elevated 1.8-10 MeV intensities. We note, however, that shortly after CME-8, at about $\sim$18:00 UT on June 1, a solar wind density increase accompanied by a magnetic field compression region arrived at STA. This compressed solar wind interval may have influenced the particle intensity enhancement on June 1 if part of particles may have been accelerated by this compression.

\end{document}